\documentstyle{article}

\newcommand{\AmS}{{\protect\the\textfont2
  A\kern-.1667em\lower.5ex\hbox{M}\kern-.125emS}}

\begin{document}
\begin{titlepage}
\pagestyle{empty}
\begin{center}
\today \hfill IPNO/TH 95-??\\
\hfill hep-lat/9508013\\

\vskip .5in

{\Large \bf
Lattice Fermions without Doublers}
\footnote{Research supported by the U.~S. Department of Energy
under contracts DE-FG04-84ER40166 and
DE-FG03-92ER40732/A004 (task B).}

\vskip .4in

Kevin Cahill\\

\vskip .15in
\baselineskip=16pt
{\em
Department of Physics and Astronomy,
        University of New Mexico\\
        Albuquerque, New Mexico 87131-1156, U.~S.~A.
\footnote{Permanent address, e-mail: kevin@cahill.phys.unm.edu}
\\
\vskip .08in
Division de Physique Th\'eorique,
\footnote{Unit\'e de Recherche des Universit\'es Paris XI
et Paris VI associ\'ee au CNRS.}
Institut de Physique Nucl\'eaire\\
91406 Orsay Cedex, France}
\end{center}
\baselineskip=26pt

\vskip 1.0in

\begin{abstract}
By placing fermions only on the even
sites of a lattice, one may halve the momentum spectrum
and construct a theory without doublers.
The interaction is nonlocal.
The fermion propagator is not a sparse matrix,
but
because the unwanted fermionic states are absent
from the formalism,
it is 256 times smaller
than the usual propagator.
\end{abstract}
\end{titlepage}

\newpage
\renewcommand{\thepage}{\arabic{page}}
\setcounter{page}{1}
\goodbreak

\section{\bf Introduction}
It is well known that the use of the natural
discretization~\cite{Wilson74} of the Dirac action leads to
16 extra states in 4 dimensions.
The standard remedies are Wilson fermions~\cite{Wilson77}
and staggered fermions~\cite{Susskind,Kogut-Susskind}.
Neither is a happy fix.
\par
The root of the doubler problem
is that the natural discretization
\begin{equation}
S_n = a^4
\sum_n
\overline \psi( n )
\left\{
- m \, \psi( n )
- \sum_\mu
{ \gamma_\mu \over 2ai }
\left[ \psi( n+\hat \mu )
- \psi( n-\hat \mu ) \right]
\right\}
\label{natdesc}
\end{equation}
of the euclidean action
approximates the derivative
by means of a gap of two lattice spacings.
In effect this factor of two doubles
the momentum spectrum and leads to
16 copies of the low-energy states.
The usual Fourier series
\begin{equation}
\widetilde{\psi} ( p ) =
{1 \over N^2 }
\sum_{n_1 = 1}^N
\sum_{n_2 = 1}^N
\sum_{n_3 = 1}^N
\sum_{n_4 = 1}^N
\exp\left[ - i { 2 \pi n \cdot p \over N } \right]
\psi( n ),
\label{utilde psi}
\end{equation}
in which $ p $ is a four-vector
of integers $ p = ( p_1, p_2, p_3, p_4 ) $
with $ 1 \le p_i \le N $,
diagonalizes the action $ S_n $
\begin{equation}
S_n = a^4
\sum_{ p_i = 1 }^N
\overline{ \widetilde{ \psi } } ( p )
\left[
- m
- \sum_{\mu = 1}^4
{ \gamma_\mu \over a }
\sin \left( { 2 \pi p_\mu \over N } \right)
\right]
\widetilde{ \psi } ( p ).
\label{daiguaction}
\end{equation}
Since $ \{ \gamma_\mu, \gamma_\nu \} = - 2 \delta_{\mu,\nu} $,
the lattice propagator is
\begin{equation}
{1 \over a^3 } \,
{ - m a + \sum_\mu \gamma_\mu \sin \left( { 2 \pi p_\mu \over N } \right)
\over
 m^2 a^2 + \sum_\mu \sin^2 \left( { 2 \pi p_\mu \over N } \right) };
\label{nprop}
\end{equation}
for $ m = 0 $
it has poles at $ p_\mu = N/2 $
and at $ p_\mu = N $
for each $ \mu $\null.
(The points $ p_\mu = 0 $
and $ p_\mu = N $ are the same.)
\par
At the price of some nonlocality,
we may leave out the unwanted states
from the start.  Thus on a lattice of even size $ N = 2 F $,
we may place independent fermionic variables
$ \psi ( 2n ) $ and $ \overline \psi( 2n ) $
only on the $ F^4 $ even sites $ 2 n $
where $ n $ is a four-vector of integers,
\begin{equation}
2 n = ( 2n_1, 2n_2, 2n_3, 2n_4 )
\label{even}
\end{equation}
and $ 1 \le n_i \le F $ for $ i = 1, .., 4 $\null.
To extend the variables $ \psi ( 2n ) $
and $ \overline \psi( 2n ) $ to the nearest-neighbor
sites $ 2n \pm \hat \mu $, we first define the Fourier variables
$\widetilde{\psi} ( k ) $ (and $ \overline{ \widetilde{\psi} } ( k ) $),
\begin{equation}
\widetilde{\psi} ( k ) =
{1 \over F^2 }
\sum_{n_1 = 1}^F
\sum_{n_2 = 1}^F
\sum_{n_3 = 1}^F
\sum_{n_4 = 1}^F
\exp\left[ - i { 2 \pi n \cdot k \over F } \right]
\psi( 2n )
\label{tilde psi}
\end{equation}
in which $ k $ is a four-vector
of integers $ k = ( k_1, k_2, k_3, k_4 ) $
with $ 1 \le k_i \le F $\null.
In terms of these Fourier variables,
the dependent, nearest-neighbor variable $ \psi ( 2n + \hat \mu ) $ is
\begin{equation}
\psi ( 2n + \hat \mu ) = { 1 \over F^2 }
\sum_{k_1 = 1}^F
\sum_{k_2 = 1}^F
\sum_{k_3 = 1}^F
\sum_{k_4 = 1}^F
\exp
\left[ i { 2 \pi \left( n + {\hat \mu \over 2 } \right)
\cdot k \over F } \right]
\widetilde{\psi} ( k ).
\label{neighbor f}
\end{equation}
In a more-compressed notation,
the dependent variable $ \psi ( 2n + \hat \mu ) $
is a Fourier series in the independent variables $ \psi ( 2m ) $
\begin{equation}
\psi ( 2n + \hat \mu ) = { 1 \over F^4 }
\sum_{ k, m }
\exp
\left[ i { 2 \pi \left( n + {\hat \mu \over 2 } - m \right)
\cdot k \over F } \right]
\psi ( 2m ).
\label{neighbor}
\end{equation}
Except in the $ \mu $ direction,
the sum over $ k_i $ gives a Kronecker delta,
and we have, with no sum over $ \mu $,
\begin{equation}
\psi ( 2n \pm \hat \mu ) = { 1 \over F }
\sum_{ l = 1 }^F \sum_{ m_\mu = 1 }^F
\exp
\left[ i { 2 \pi \left( n_\mu \pm {1 \over 2 } - m_\mu \right)
l \over F } \right]
\psi ( 2n + ( 2 m_\mu - 2 n_\mu ) \hat \mu ).
\label{neighbor mu}
\end{equation}
\par
Let us define the lattice delta functions
\begin{equation}
\delta^\pm( 2 j ) = { 1 \over F }
\sum_{ l = 1 }^F
\exp
\left[ i { 2 \pi \left( \pm {1 \over 2 } - j \right)
l \over F } \right].
\label{delta}
\end{equation}
Then the nearest-neighbor variables may be expressed as
the sums
\begin{equation}
\psi ( 2n \pm \hat \mu ) =
\sum_{j = 1 - n_\mu }^{ F - n_\mu  }
\delta^\pm( 2 j ) \psi ( 2n + 2j \hat \mu ).
\label{nn}
\end{equation}
The action with independent fields at only $ F^4 $
sites is
\begin{equation}
S = (2a)^4
\sum_{n_i = 1}^F
\overline{ \psi } ( 2n )
\left\{
- m \, \psi( 2n )
- \sum_{\mu = 1}^4
{ \gamma_\mu \over 2ai }
\left[
\psi( 2n + \hat \mu ) - \psi( 2n - \hat \mu )
\right]
\right\},
\label{action}
\end{equation}
in which the sum is over $ 1 \le n_i \le F $
as in eq.(\ref{tilde psi}).
In terms of the lattice
delta functions (\ref{delta}),
this action is
\begin{eqnarray}
S & = & (2a)^4
\sum_{n_i = 1}^F
\overline{ \psi } ( 2n )
\big\{
- m \, \psi( 2n )
\nonumber \\
& & \mbox{}
- \sum_{\mu = 1}^4
{ \gamma_\mu \over 2ai }
\sum_{j = 1 - n_\mu}^{F - n_\mu}
\left[
\delta^+ ( 2j ) - \delta^- ( 2j )
\right]
\psi( 2n + 2j \hat \mu )
\big\}.
\label{daction}
\end{eqnarray}
\par
\par
We may now verify that there are no doublers.
The Fourier series (\ref{tilde psi}) and (\ref{neighbor f})
diagonalize the action (\ref{daction})
\begin{equation}
S = (2a)^4
\sum_{ k_i = 1 }^F
\overline{ \widetilde{ \psi } } ( k )
\left[
- m
- \sum_{\mu = 1}^4
{ \gamma_\mu \over a }
\sin \left( { \pi k_\mu \over F } \right)
\right]
\widetilde{ \psi } ( k )
\label{daigaction}
\end{equation}
in which all the sums over $ k_i $ run
from 1 to $ F $\null.
Now the lattice propagator is
\begin{equation}
{1 \over a^3 } \,
{ - m a + \sum_\mu \gamma_\mu \sin \left( { \pi k_\mu \over F } \right)
\over
 m^2 a^2 + \sum_\mu \sin^2 \left( { \pi k_\mu \over F } \right) }.
\label{prop}
\end{equation}
For $ m = 0 $ this propagator
has poles only at $ k_\mu = F $,
which is the same point as $ k_\mu = 0 $\null.
There are no doublers.
\par
Gauge fields may be
added in the usual way.
In a gauge theory with a gauge field $ U_\mu( n ) $
defined on the link $ ( n, n + \hat \mu ) $,
one may construct
the ordered product
$ U( 2n, 2n + 2j \hat \mu ) $
of Wilson links $ U_\mu( n ) $
from site $ 2n + 2j \hat \mu $ to site $ 2n $
for $ j > 0 $ as
\begin{equation}
U( 2n, 2n + 2j \hat \mu ) =
U_\mu( 2n ) U_\mu( 2n +  \hat \mu ) \dots
U_\mu( 2n + ( 2j - 1 ) \hat \mu )
\label{U+}
\end{equation}
and for $ j < 0 $ as
\begin{equation}
U( 2n, 2n + 2j \hat \mu ) =
U^\dagger_\mu( 2n - \hat \mu ) \dots
U^\dagger_\mu( 2n - 2|j| \hat \mu ).
\label{U-}
\end{equation}
Thus one may covariantize
the action (\ref{daction})
by inserting the line $ U( 2n, 2n + 2j \hat \mu ) $
\begin{eqnarray}
\lefteqn{S = (2a)^4
\sum_{n_i = 1}^F
\overline{ \psi } ( 2n )
\big\{
- m \, \psi( 2n ) }
\nonumber \\
& & \mbox{}
- \sum_{\mu = 1}^4
{ \gamma_\mu \over 2ai }
\sum_{j = 1 - n_\mu}^{F - n_\mu}
\left[
\delta^+ ( 2j ) - \delta^- ( 2j )
\right]
U( 2n, 2n + 2j \hat \mu )
\psi( 2n + 2j \hat \mu )
\big\}.
\label{kaction}
\end{eqnarray}
\par
Because of the lack of locality,
the fermion propagator is not a sparse matrix.
On the other hand,
there are only one-sixteenth as many
fermionic variables
and so the fermion propagator
is smaller by a factor of 256.
The present formalism
of thinned fermions
therefore may be useful
in practise as well as in principle.

\section*{\bf Acknowledgements}
I should like to thank M.~Creutz, S.~Nicolis, J.~Smit, and J.~Stern
for helpful conversations
and A.~Comtet and D.~Vautherin
for inviting me to Orsay.
I am grateful to the Department of Energy for financial support
under grant DE-FG04-84ER40166
and task B of grant DE-FG03-92ER40732/A004.


\begin{thebibliography}{99}
\bibitem{Wilson74} K.~Wilson,
{\sl Phys.~Rev.\ \/} D10 (1974) 2445.
\bibitem{Wilson77} K.~Wilson, in {\sl New Phenomena
in Subnuclear Physics, Part A\/}, A.~Zichichi, ed.\
(Plenum, New York, 1977), p.~69.
\bibitem{Susskind} L.~Susskind,
{\sl Phys.~Rev.\/} D16 (1977) 3031.
\bibitem{Kogut-Susskind} T.~Banks, S.~Raby, L.~Susskind,
J.~Kogut, D.~R.~T. Jones, P.~Scharbach, and D.~Sinclair,
{\sl Phys.~Rev.\/} D15 (1977) 1111.
\end{thebibliography}
\end{document}